# Demonstrating magnetic memory in iron-rhodium structures using a quantum diamond microscope

Kristine V. Ung[1, 2, 3, *], Gregory M. Stephen[4], Nicholas A. Blumenschein[4], Alexander J. Edwards[4, 6], Samuel W. LaGasse[4], Steven P. Bennett[5], Aubrey T. Hanbicki[4], Ronald L. Walsworth[1, 2, 3, 6], Adam L. Friedman[4], Paul V. Petruzzi[4]

[1]*Joint Quantum Institute, University of Maryland, College Park, MD 20742, USA*
[2]*Department of Physics, University of Maryland, College Park, MD 20742, USA*
[3]*Quantum Technology Center, University of Maryland, College Park, MD 20742, USA*
[4]*Laboratory for Physical Sciences, 8050 Greenmead Dr., College Park, MD 20740, USA*
[5]*The U.S. Naval Research Laboratory, 4555 Overlook Ave. SW, Washington, DC, 20375, USA*
[6]*Department of Electrical and Computer Engineering, University of Maryland, College Park, MD 20742, USA*

## Abstract

Iron-rhodium (FeRh) has a first-order phase transition near room temperature between antiferromagnetic (AFM) and ferromagnetic (FM) phases, making it a promising material for magnetic memory technologies like heat-assisted magnetic recording (HAMR). It has a comparatively sharper phase transition and lower writing temperature than alternative materials, implying less thermal engineering constraints and an increase in write/read head lifetime. Despite great effort, however, AFM-based magnetic memory using FeRh has not yet been realized. Here, we employ both wide-field and scanning nanoscale quantum diamond microscopes (QDMs) to image directly the magnetic field of a patterned FeRh thin film structure under ambient conditions, demonstrating a magnetic recording technique that is reliable and robust. We experimentally identify coupling between the Néel and magnetization vector directions; and also, that the magnetic orientation of the FM phase uniquely determines the Néel vector in the AFM phase, due to pinned uncompensated magnetic moments (UMMs) in the FeRh structure. Thus, the magnetic orientation is maintained when the system is cycled between AFM and FM phases, providing the foundation for a practical, AFM-based magnetic memory.

## Introduction

The rise of cloud computing and AI data centers has generated an increasing demand for low-cost, high-density storage. Heat-assisted magnetic recording (HAMR) overcomes the thermal stability limits of conventional magnetic hard disc drives (HDDs) by using high-coercivity materials, such as $L1_0$-FePt, as the storage medium. However, the magnetic domains must be heated near their Curie temperature (~750 K) to reduce the magnetic coercivity sufficiently for an HDD write head to control their magnetization [1]. This heating requires significant localized energy from a near-field transducer to reach the required temperature, degrading the read/write heads of the HAMR drive with repeated use — the dominant failure mechanism of the drive [2], [3], [4]. Using a material that requires a lower write temperature will reduce the engineering complexity of HAMR drives and extend their lifetime by easing thermal design constraints and potentially eliminating material accumulation on the head surface [5], [6].

Iron-rhodium (FeRh) is a binary intermetallic compound that has been studied for over 60 years due to its unique metamagnetic phase transition near room temperature [7]. FeRh is a G-type antiferromagnet at room temperature with magnetic moments on the Fe sites and no moment on the Rh site. FeRh transitions to the ferromagnetic (FM) phase slightly above room temperature, where the Rh site gains a small magnetic moment that is aligned with the Fe sites. This transition from antiferromagnetic (AFM) to FM ordering results in a large change in magnetization (~800 $emu/cm^3$) and a 1.0% increase in lattice constant [8], [9]. The transition temperature, normally around 450 K, decreases with substitutional doping [10], strain [11], [12], [13], [14], [15] lithographic patterning [16], [17] [18], [19], [20] , and ion implantation [14], [21], [19]. The properties of FeRh make it a desirable material for commercial spintronic logic, memory devices and, as demonstrated in this work, HAMR drives. In addition to the lifetime increase already discussed, the first-order phase transition of FeRh leads to a sharper change in magnetic coercivity than that of the second-order FM-paramagnetic transition of alternative materials such as FePt, meaning each bit should be less susceptible to magnetic instability, potentially reducing back-switching of the bits and decreasing read errors [22].

Extensive research has been dedicated to elucidating the physics of AFM-based memory with different device geometries and materials coupled with FeRh. Most efforts attempt to create magnetic orderings of FeRh in contact with lateral and vertical heterostructures of different materials [23], [24], [25], [26]. Additional attempts to use FeRh for magnetic memory rely on either rotating the Néel vector [27], or using the magnetic phase itself as the state variable [28]. Despite longstanding efforts, however, a functional approach for AFM-based memory has not yet been realized.

In this work, we demonstrate coupling between the Néel and FM vector directions in an FeRh thin film structure near room temperature, and importantly, show that the FM state can be reliably read out when in the AFM phase, even after repeated cycling between phases. Consistent with the nominal physical picture of the FM phase, we find that rotating the FM direction rotates the Néel vector. Less intuitively, given the collinear antiferromagnetic symmetry of the Fe moments in the AFM phase, we also observe deterministic switching of the AFM phase by simply flipping the FM direction using a modest applied 'poling' magnetic field. By imaging the stray magnetic field from the FeRh structure using both wide-field and scanning nanoscale quantum diamond microscopes (QDMs), we characterize the key physical mechanism behind this FM-to-AFM magnetic memory: poling the FeRh magnetization in the FM phase sets the Néel vector direction, which can then be read out in the AFM phase due to pinned uncompensated magnetic moments (UMMs) in the FeRh structure. This surprising result demonstrates that FeRh retains its previously poled FM state after becoming AFM, providing a promising path to a practical, AFM-based magnetic memory.

# Methods

**FeRh Structure Fabrication**

Magnetron sputtering is used to deposit 35 nm of FeRh onto an MgO substrate. The film is etched into rectangular structures with varying dimensions using an ion mill. Additional details about the FeRh film growth, crystalline quality, and device fabrication have been previously reported elsewhere [28]. The results shown in this work are acquired from a 25 μm × 100 μm FeRh structure, as it has a clear easy axis and its magnetic behavior is easily resolvable by the wide-field and scanning nanoscale QDMs.

**NV magnetometry**

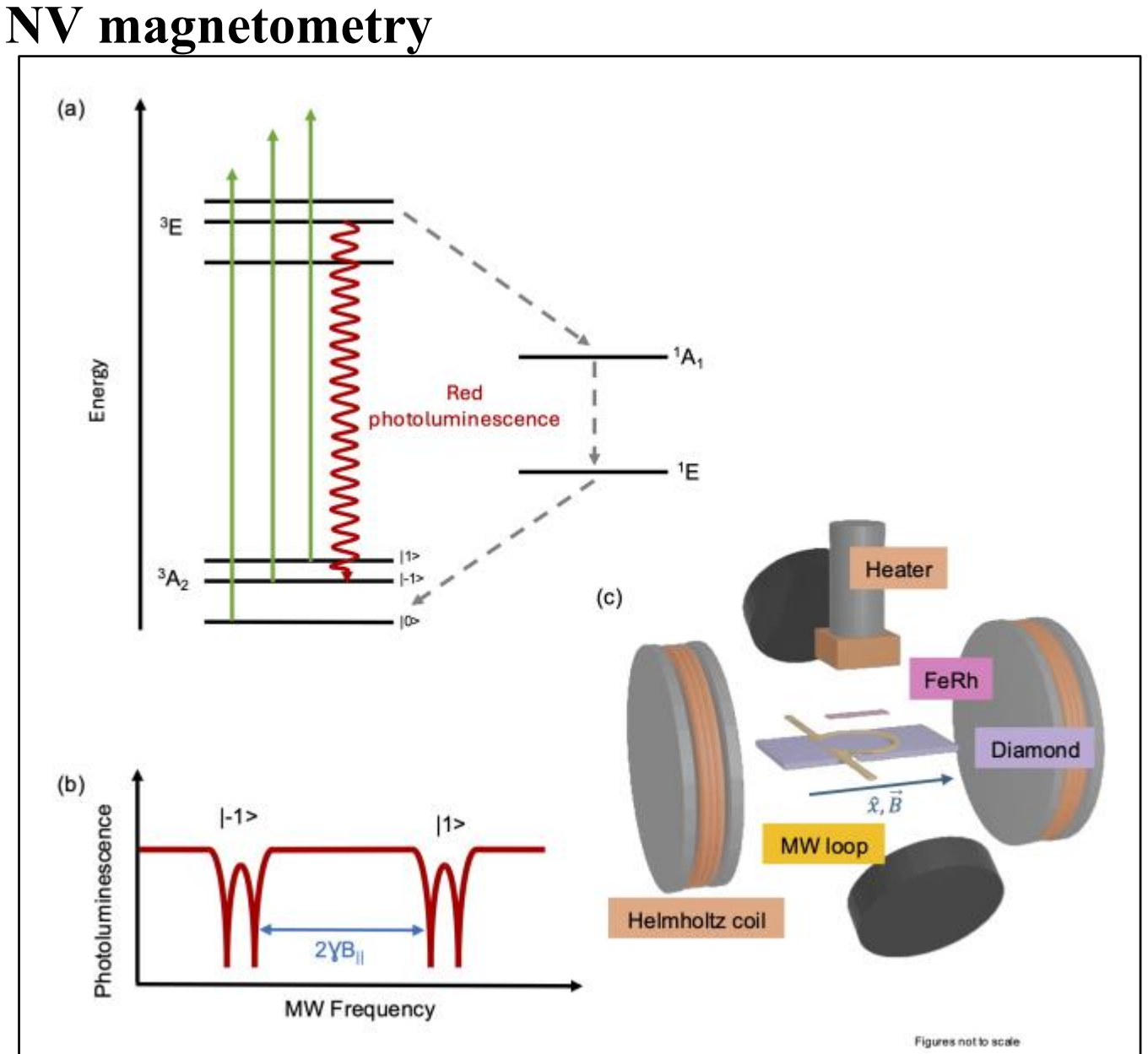

*FIG 1. (a) Energy level diagram of the negatively charged nitrogen vacancy (NV) center. (b) NV photoluminescence (PL) as a function of applied microwave signal frequency, where the PL dips represent NV spin resonances that coincide with the local magnetic field strength, known as optically detected magnetic resonance (ODMR).* ***(c)*** *Schematic drawing of the wide-field quantum diamond microscope (QDM) experimental setup used for magnetic imaging of the FeRh film structure.*

The stray magnetic field pattern of the FeRh structure is first measured (imaged) by the wide-field QDM while in the AFM and FM phases to demonstrate FM magnetization and AFM Néel vector coupling. The wide-field QDM is a well-established tool for quantitatively mapping magnetic fields in diverse samples at the micron-scale [20]. Our wide-field QDM employs a CVD-grown diamond chip (4 mm x 2 mm), 99.99% $^{12}$C, with a 10 $\mu m$ thick, $^{15}$N-enriched surface layer ([N] = 17 ppm). Postgrowth treatment via electron irradiation and annealing creates a 2.7 ppm concentration of negatively charged nitrogen vacancy (NV) centers in the surface layer. The wide-field QDM provides magnetic images with a field-of-view of $235\mu m \times 245\ \mu m$, spatial resolution of $10\ \mu m$, and sensitivity to static magnetic fields of about $300\ \mathrm{nT}/\sqrt{\mathrm{Hz}}$ in each binned $10\ \mu m$ pixel, allowing FeRh feature measurements to be completed in minutes. NV magnetometers are also robust against temperature changes, which is critical to quantitatively compare the magnetic field of the FeRh structure in the AFM and FM phases.

Negatively charged NV centers are solid-state defects in diamond with electronic spin triplet ground and excited states that possess a spin-sensitive photoluminescence (PL) amplitude allowing for preparation and optical readout of the spin sublevels $m_s = |0\rangle, |{+1}\rangle$, and $|-1\rangle$, where $m_s$ is the spin projection along the NV center quantization axis. The ground state spin sublevels have a zero-field splitting of 2.87 GHz, allowing coherent control of the spin state using microwave fields. The application of a DC magnetic field splits the degenerate $|\pm 1\rangle$ spin states by a frequency equal to $f = \gamma B$ where $\gamma$ is the electron gyromagnetic ratio of 28 GHz/T, and $B$ is the magnetic field magnitude along the axis of the defect. Including the zero-field splitting, $D$, the transition (resonance) frequencies from the $|0\rangle$ to the $|\pm 1\rangle$ states are:

$$f_{\pm} = D \pm \gamma B \qquad (1)$$

where $\pm$ indicates transitions to the $|+1\rangle$ and $|-1\rangle$ states, respectively [29].

A continuous-wave optically detected magnetic resonance (CW-ODMR) sensing protocol is used to measure the magnetic field from the FeRh structures. By sweeping the microwave (MW) frequency around $f_{\pm}$, the CW-ODMR protocol allows accurate measurement of the NV spin resonance frequencies, as shown in **Fig. 1b**. [29], [30], [31], [32], [33]. Note that the double Lorentzian shape in **Fig. 1b** is due to the NV electronic spin hyperfine interaction with the $^{15}$N nucleus (I=1/2). Fitting of the per-pixel ODMR spectra in a QDM magnetic image is performed using an open-source software package designed for NV experiments [34], allowing extraction of the NV spin resonance frequencies $f_+$ and $f_-$ in **Equation 1.** Finally, the per=pixel magnetic field $B$ along the NV axis is determined from:

$$B = \frac{f_+ - f_-}{2\gamma} \qquad (2)$$

**Fig. 1c** shows a schematic of the wide-field QDM experimental setup, where the FeRh sample is mounted onto a copper block fastened to a cartridge heater, allowing for sample temperature control; and a thermocouple attached to the copper block measures the sample temperature. The FeRh sample is suspended less than 10 microns above the diamond and closest to the NV sensing layer. A Helmholtz coil generates an external magnetic poling field of about $\pm 10$ mT along the easy axis of the FeRh structure. Further information regarding the diamond composition, experimental equipment, and background field subtraction is detailed in the Supplementary Information (SI).

## Results/Discussion

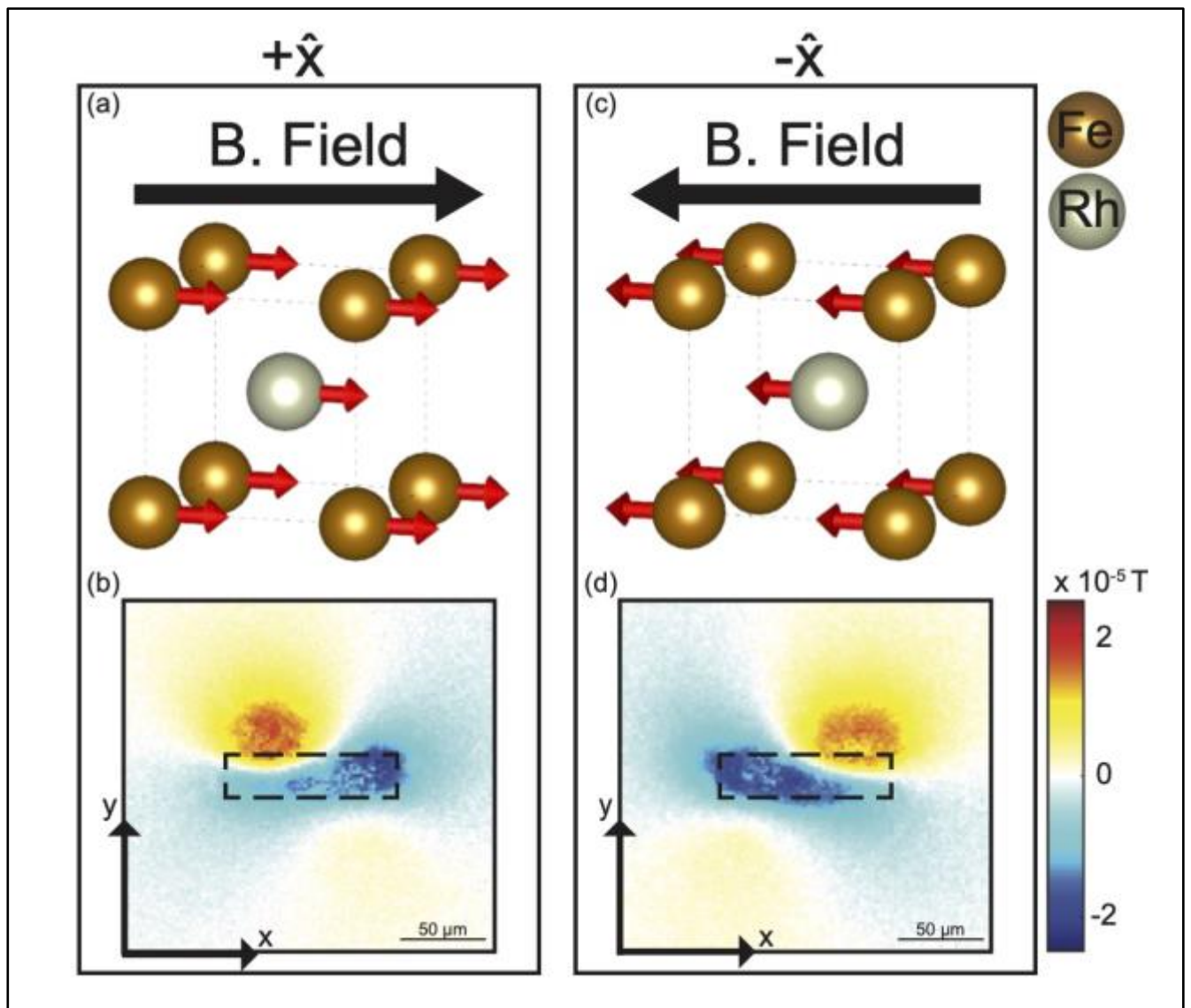

*Fig.2**(a)** Magnetization direction and **(b)** wide-field QDM magnetic field map of the FeRh structure when poled along $+\hat{x}$. **(c)** Magnetization direction and **(d)** wide-field QDM magnetic field map of the FeRh structure when poled along $-\hat{x}$. The black dashed boxes in **(b)** and **(d)** represent the position of the FeRh*

**Fig. 2(a, c)** illustrates the magnetic moment arrangement of the FeRh unit cell in the FM phase. The shape anisotropy created by the 4:1 aspect ratio of the structure defines the magnetic easy axis as the $x$-direction and ensures only two stable magnetic orientations. Application of the magnetic poling field along the $\pm\hat{x}$-direction nominally causes the sample magnetization to point in the same direction. For the rest of this manuscript, we refer to the orientation after an external field is applied to the FM phase in the $\pm\hat{x}$-direction as the $\pm\hat{x}$ state.

The direction of the magnetization of the FeRh structure is determined by wide-field QDM magnetic imaging. **Fig. 2(b, d)** shows that poling in the $\pm\hat{x}$-direction results in magnetic field images that are mirrored around the y-axis, with the color bar representing the strength of the magnetic field vector along the NV sensing axis. The wide-field QDM maps display magnetic field projections along the (110) diamond crystal axis (*i.e.,* along the quantization axis of the probed NV ensemble), which, relative to the FeRh structure, is along the y-axis in the x-y plane and 54.7° from the normal. As a result, the observed magnetic field pattern is a weighted superposition of the y and z components of the stray magnetic field from the FeRh structure. The wide-field QDM magnetic field maps are indicative of the direction of magnetization set by the poling field, as confirmed by magnetic field calculations (see SI for additional information). Therefore, we can use distinct magnetic field maps to identify the direction of the FeRh magnetization.

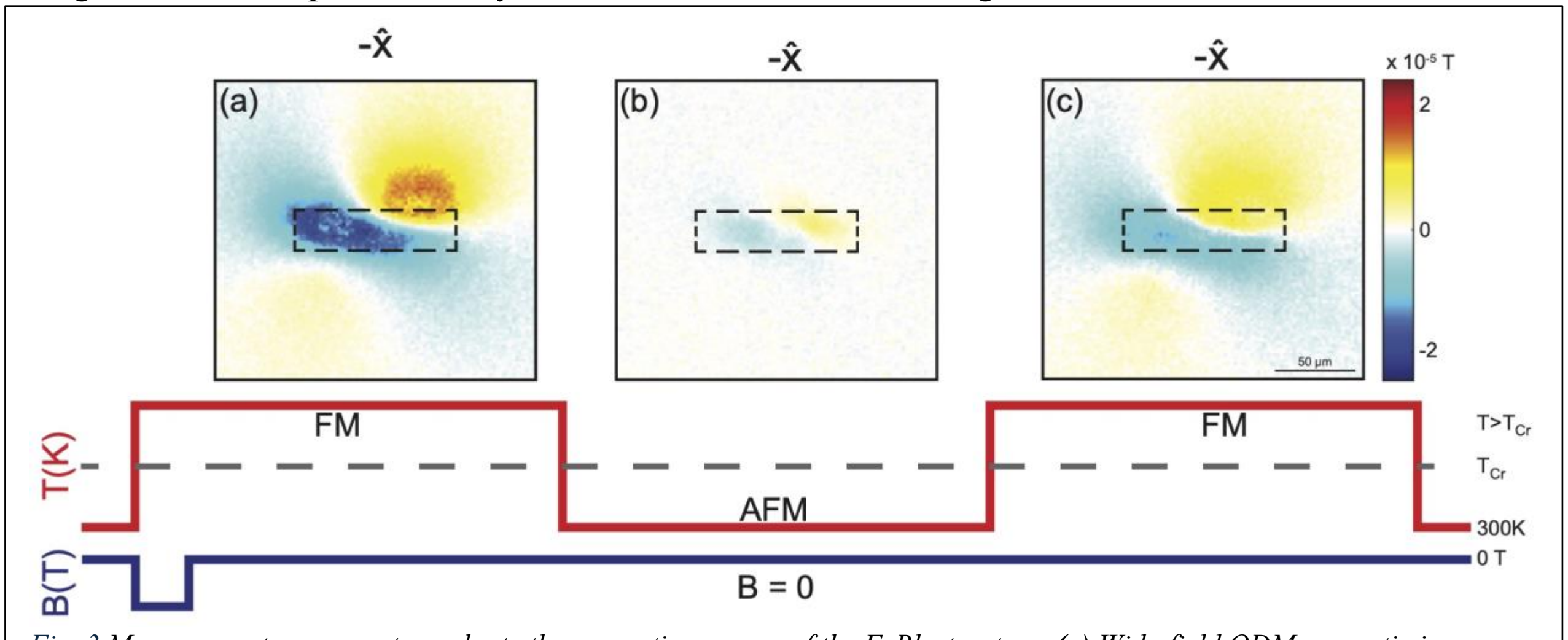

*Fig. 3 Measurement sequence to evaluate the magnetic memory of the FeRh structure.* ***(a)*** *Wide-field QDM magnetic image taken in the FM phase at temperature ≈ 450 K, after the sample is poled in the* $-\hat{x}$ *direction.* ***(b)*** *Room temperature wide-field QDM magnetic image in the AFM phase.* ***(c)*** *Wide-field QDM magnetic image in the FM phase with no prior poling.*

Because the AFM phase has an equal number of magnetic moments in either direction, we expect that there is no preferred direction and the magnetization orientation after the FM transition is determined stochastically. To assess whether the magnetization vector is truly stochastic when transitioning from AFM-to-FM phase, we thermally cycle between these two phases and then image the magnetic field from the FeRh structure. If the magnetization direction after the transition is random, we expect to measure the $\pm\hat{x}$ states with equal probability. To test this behavior, we perform a memory cycle measurement sequence consisting of three wide-field QDM magnetic images, as shown in **Fig. 3**. The red line represents the temperature of the FeRh structure, and the dashed grey line represents the critical temperature where FeRh exhibits its AFM-FM phase transition. Above the critical temperature (≈ 450 K), FeRh is in the FM phase; and at room temperature (≈300 K) it is in the AFM phase. The blue line represents the external (poling) magnetic field generated by the Helmholtz coil. When the blue line is greater (less) than zero, an applied field is present in the $+\hat{x}$ ($-\hat{x}$) direction. The measurement cycle starts with heating the FeRh structure into the FM phase and poling it along $-\hat{x}$, as confirmed by the wide-field QDM magnetic field map (**Fig. 3a**). The sample is then cooled down to the AFM phase, and the magnetic field is imaged again **(Fig. 3b)**. A small residual field is observed in the AFM phase stemming from defects in the FeRh film, as discussed at the end of this section. Finally, the FeRh structure is heated back into the FM phase, this time without poling, and the magnetic field is imaged a third time **(Fig. 3c)**. From this final magnetic field map, we determine that the FeRh structure is in the $-\hat{x}$ state, the same as when it was poled initially (**Fig. 3a**). This measurement cycle is then repeated except with FM poling along $+\hat{x}$: we find that the FeRh structure returns to the $+\hat{x}$ state after temperature cycling (*i.e.,* FM-to-AFM-to-FM). The three-measurement memory cycle is repeated over 20 times for both $-\hat{x}$ and $+\hat{x}$ poling, with essentially identical magnetic maps observed. These results show that the FeRh structure returns to its previously poled state after the FM-to-AFM-to-FM cycle is concluded, demonstrating that the orientation of both the FM moment and the Néel vector are coupled, creating a magnetic memory effect where the FeRh structure retains its previously poled state.

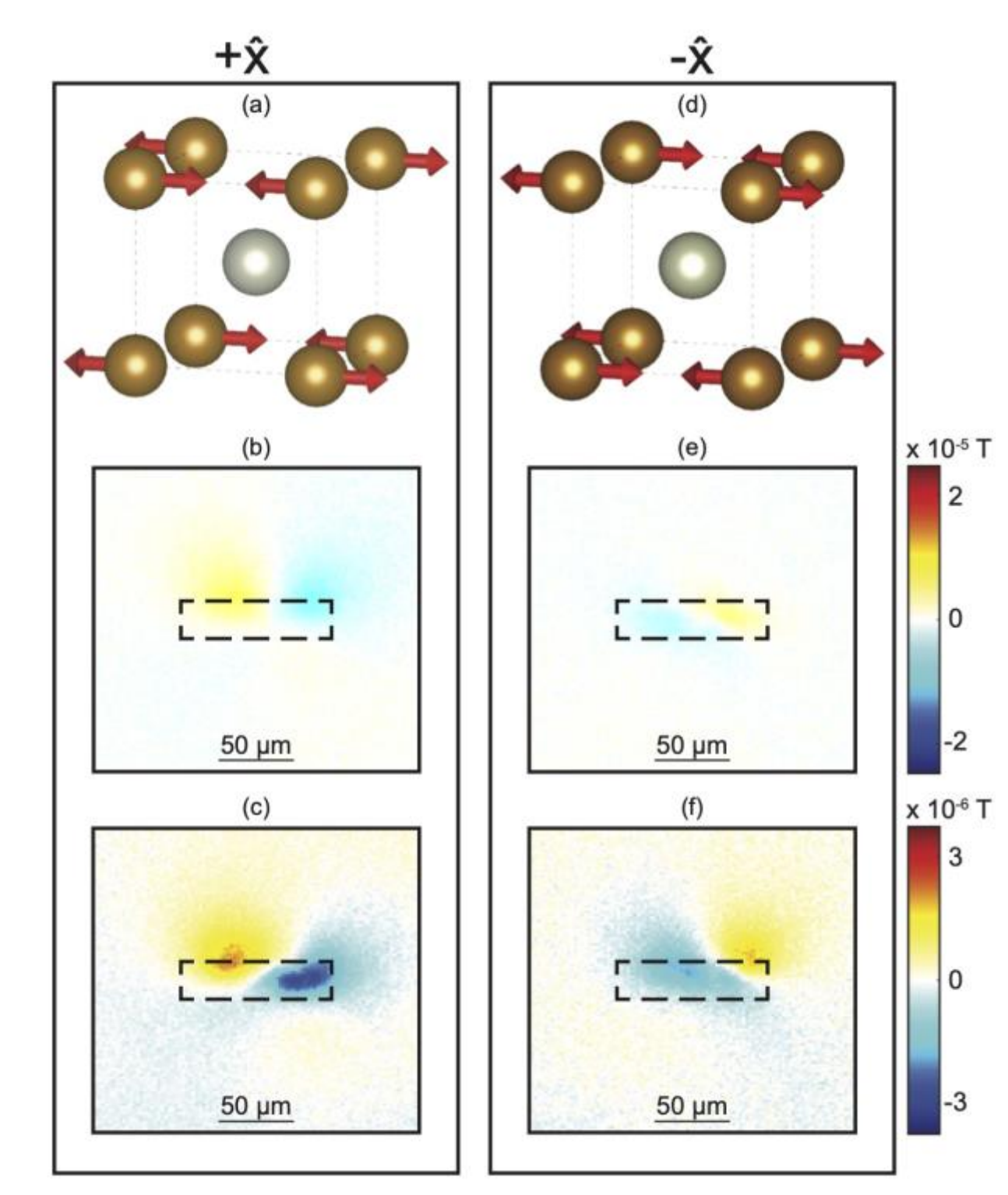

*Fig 4.* ***(a)*** *AFM ordering of FeRh after being poled in state $+\hat{x}$, representing the Néel vector in that direction.* ***(b)*** *Wide-field QDM magnetic field map of FeRh structure in the AFM phase after initially being poled in state $+\hat{x}$ in the FM phase.* ***(c)*** *Magnetic field map resulting from pinned uncompensated magnetic moments (UMMs) in the FeRh structure after initially being poled in state $+\hat{x}$.* ***(d)*** *AFM ordering of FeRh after being poled in state $-\hat{x}$, representing the Néel vector in that direction.* ***(e)*** *Wide-field QDM magnetic field map of FeRh structure in the AFM phase after initially being poled in state $-\hat{x}$.* ***(f)*** *Magnetic field map resulting from pinned UMMs after initially being poled in state $-\hat{x}$ in the FM phase.*

As noted above and demonstrated in **Fig. 3,** when the FeRh structure is in the AFM phase, wide-field QDM magnetic images reveal a residual field about 2.5 times smaller than in the FM phase; see also **Fig. 4b** and **Fig. 4e** for wide-field QDM magnetic images in the AFM phase after poling the structure in the $+\hat{x}$ and the $-\hat{x}$ directions while in the FM phase, respectively. We use these distinct magnetic field maps to determine the direction of the Néel vector and thereby identify the AFM states as $+\hat{x}$ and the $-\hat{x}$ as well. Nominally, the two possible FeRh AFM orderings (**Fig. 4a, d**) result in a net zero magnetization for the structure. However, unpinned and pinned uncompensated magnetic moments (UMMs) exist in antiferromagnetic materials [35], [36], including FeRh [37], resulting in a non-zero net magnetic moment with a measurable magnetic field in the AFM phase [38], [39], [40]. When the material is cooled to the AFM phase, UMMs are pinned by the AFM ordering and create a unidirectional anisotropy [35]. The stray magnetic field from these pinned UMMs make the relationship between the AFM phase magnetization and the orientation (along $\pm\hat{x}$) of the Néel vector evident in our measurements, as we find distinctly different magnetic field maps for $-\hat{x}$ and $+\hat{x}$ in the AFM phase.

To examine the underlying physical phenomena of the observed magnetic memory in the FeRh structure, we acquire a magnetic field map of the AFM phase in the $+\hat{x}$ state, *i.e.,* after initially being poled in state $+\hat{x}$ in the FM phase (**Fig. 4b**). Then, while the structure is in the AFM phase, we pole the structure in the $-\hat{x}$ direction, acquire a second magnetic field map, and add the two images pixel by pixel as shown in **Fig. 4c**. Since the unpinned UMMs follow the poling field, they cancel when the two images are added, leaving only the magnetic field map resulting from the pinned UMMs [39]. In **Fig. 4c** we see that the magnetic image of the pinned UMMs in the AFM phase is similar in spatial structure to that of the FM $+\hat{x}$ state, albeit with a reduced field magnitude in comparison to **Fig. 2b**. The same wide-field QDM magnetic images and post processing in the AFM phase are performed after the FeRh structure is initially poled in state $-\hat{x}$ in the FM phase (**Fig. 4e**). The results for pinned UMMs in the AFM phase (**Fig. 4f**) are found (again) to be similar to that of the FM $-\hat{x}$ state with about an order of magnitude smaller field magnitude (**Fig. 2d**).

These QDM magnetic images of the AFM phase are strong evidence that the memory effect stems from pinned UMMs in the FeRh structure; and that the direction of the Néel vector is coupled to the FM magnetization vector. When transitioning between FM and AFM phases, these UMMs act as a poling field to set the magnetization vector. Results from other groups have shown that the ratio of the magnetization produced by the UMMs to the magnetization when the FeRh is in the FM phase varies with Fe/Rh composition, growth parameters, and film thickness [38]. The ratio of the maximum magnetic field of pinned UMMs to the maximum magnetic field in the FM phase for the FeRh structure measured here is similar to previously published results [38], [40].

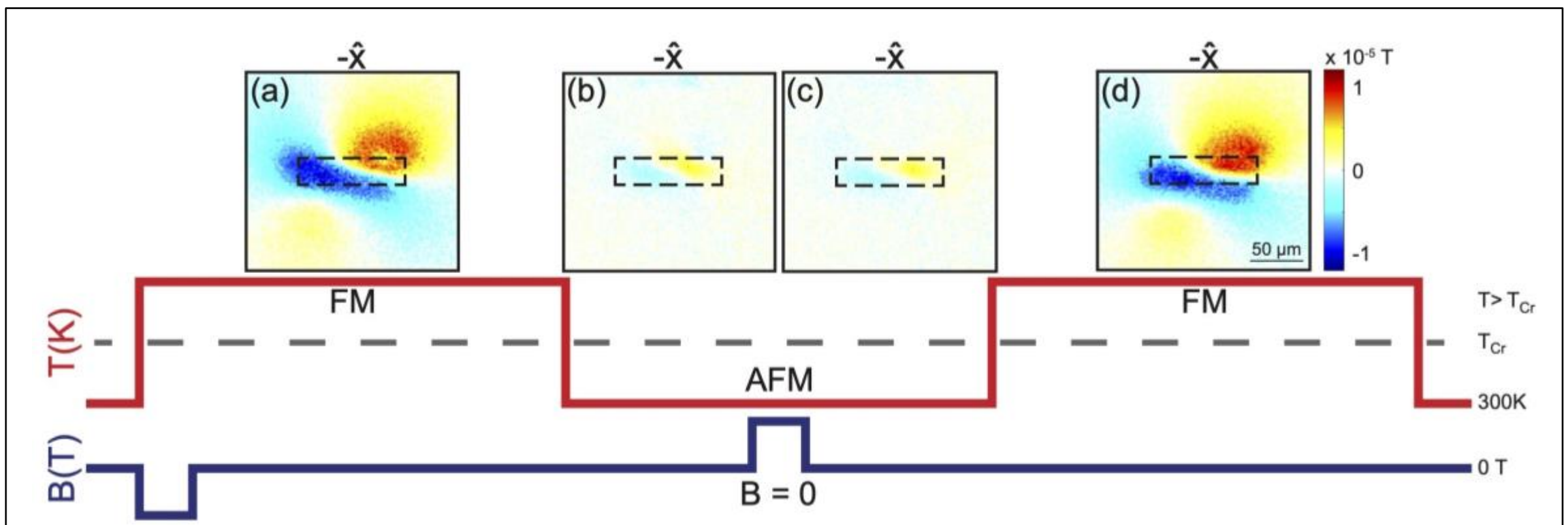

*Fig. 5. Modified measurement sequence to test the robustness of the FeRh structure magnetic memory, with an additional poling field applied in the AFM phase.* ***(a)*** *Wide-field QDM magnetic image taken at temperature ≈ 450 K, after the sample is poled to state $-\hat{x}$ in the FM phase.* ***(b)*** *Room temperature wide-field QDM magnetic image of the sample in the AFM phase.* ***(c)*** *Nominal change observed in a room temperature wide-field QDM magnetic image after an attempt to pole the FeRh structure to state $+\hat{x}$ in the AFM phase.* ***(d)*** *Wide-field QDM magnetic image taken after return to the FM phase is similar to image in (a), indicating the sample remains in state $-\hat{x}$ and thus memory is conserved.*

Additionally, we test the robustness of the magnetic memory by poling the FeRh structure at room temperature. As shown in **Fig. 5**, we first put the sample into the FM $-\hat{x}$ state and a wide-field QDM magnetic image is acquired **(Fig. 5a)**. Then, the sample is cooled into the AFM phase, and the residual field is again imaged **(Fig. 5b)**. While the FeRh structure is still in the AFM phase, an external field is applied to attempt to flip the sample into the $+\hat{x}$ state, followed by another wide-field QDM magnetic image (**Fig. 5c**). We see that the applied field does not flip the pinned UMMs and the sample remains in the $-\hat{x}$ state when returning to the FM phase **(Fig. 5d)**. Similar results are found for QDM measurements taken with the sample starting in the FM $+\hat{x}$ state, and then an external field applied in the $-\hat{x}$ direction in the AFM phase. We therefore conclude that an applied field strong enough to flip the magnetization in the FM phase is not large enough to flip the magnetization of the pinned UMMs, because they are now pinned to the AFM ordering.

To understand where these UMMs reside in the FeRh thin film, we perform magnetic imaging with a commercially available scanning nanoscale QDM (see SI). This instrument harnesses the same NV physics as the wide-field QDM, but with higher spatial resolution and reduced field-of-view, as it utilizes a single NV on an atomic force microscope tip. **Fig. 6a** shows an optical image of the FeRh thin film. We focus on a small region near the edge and image the pinned UMMs as shown in **Fig. 6b**, with similar post-processing of magnetic images as that described above. From these measurements, we observe a relatively higher density of pinned UMMs near the edge of the FeRh structure compared to in the bulk [39]. These edge features are likely due to dangling bonds created during the ion milling process. This high density of UMMs along the edge are of interest for future studies, e.g., how the ratio of bulk to edge states affects the overall memory characteristics.

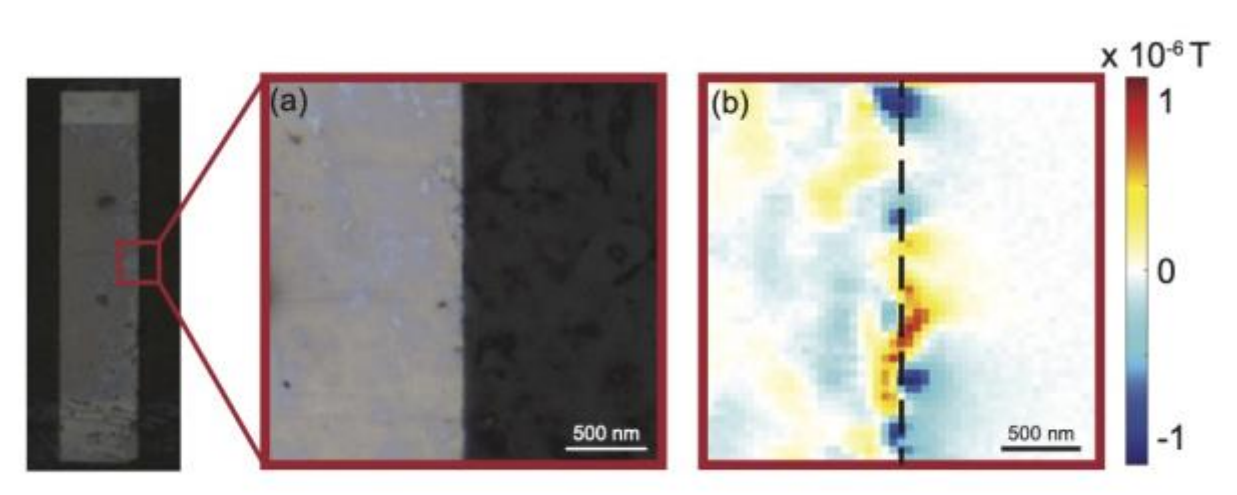

*Fig. 6.* ***(a)*** *High-magnification optical image of FeRh thin film edge.* ***(b)*** *Magnetic field map of pinned magnetic moments in the FeRh structure with the same field of view as (a), acquired with a scanning nanoscale QDM. The black dashed line represents the edge of the FeRh structure.*

To support the claims that (*i*) UMMs within the material are the primary cause of the observed residual field pattern in the AFM phase, and (*ii*) the UMM magnetization is strongly pinned to the AFM lattice, we simulate a hybrid FM/AFM system in mumax$^+$ [41], [42]. The simulation represents a single UMM in the AFM system as follows: a 64 × 64 × 64 nm$^3$ FeRh antiferromagnet cube containing a 32 nm diameter sphere within which antiferromagnetic FeRh is replaced with ferromagnetic FeRh. AFM spins are frozen to simulate boundary conditions from a much larger AFM lattice surrounding the UMM site. Exchange energies are used to control the interactions within the system such that the AFM lattice can strongly reinforce the magnetization of the FM sphere, able in the best case to withstand fields greater than 100 T. The FM magnetization does not precess, with z-directed magnetization equal to $M_{sat}$ (± simulation precision) for the duration of the simulation. In contrast, when the FM sphere is simulated alone with no surrounding AFM system, the magnetization precesses as expected, indicating that the presence of the AFM region pins the FM magnetization. The external field is then increased in the simulation for both the FM and AFM regions, and a slight x-component of magnetization develops, enabling calculation of the effective z-directed field induced upon the FM sphere by the surrounding AFM region, with $B_{eff,z} = B_{ext,x} * m_z/m_x$. $B_{eff,z}$ is greater than 108 T for all external fields tested; also, $B_{eff,z}$ increases as the magnetization is pushed away from exchange equilibrium. Additional simulation details are included in the SI. Note that these simulated values for $B_{eff,z}$ are well above the experimentally observed magnetic field required to force the AFM-FM transition near 4 K [43], indicating strong interatomic pinning.

# Conclusion

Through quantum diamond microscope (QDM) magnetic field imaging, we show that pinned uncompensated magnetic moments (UMMs) in a FeRh thin film structure near room temperature follow the FeRh AFM order and do not flip direction. This effect is confirmed experimentally for an applied magnetic field of at least 10 mT, though we expect UMM pinning to persist to much higher magnetic fields due the high coercivity of the AFM phase. The combination of low FM phase coercivity, strong AFM phase coercivity, the net magnetic moment produced by pinned UMMs near 300 K, and the relatively low phase transition temperature, make FeRh an attractive candidate for non-volatile, AFM-based magnetic memory applications. For example, FeRh could be a drop-in replacement in heat-assisted magnetic recording (HAMR), which may decrease the overall thermal engineering constraints and increase the stability of the magnetic media.

This research provides the foundation for further studies to explore FeRh utilization in HAMR devices. Future work is needed to determine optimal patterning strategies to maximize device performance and storage density. Studies that provide information on FeRh shape anisotropy limitations, methods to create out-of-plane magnetization, and general patterning strategies to minimize individual domain spacing are critical steps in developing a feasible FeRh HAMR device with increased memory per unit area compared to conventional nanoparticle-based devices.

## Acknowledgements

The authors gratefully acknowledge the helpful scientific discussions on NVs from those in the pulsed-imaging team in the Walsworth group and the critical assistance from LPS support staff including G. Latini, J. Wood, R. Brun, P. Davis, D. Crouse, and D. Basu through the course of this project. The material in this manuscript is based on research by K.V.U supported by a Joint Quantum Institute (JQI) graduate research fellowship. This work is supported by the Maryland Procurement Office under Award No. H9823019C0220; and the University of Maryland Quantum Technology Center.